\documentclass[%
 reprint,
superscriptaddress,
 amsmath,amssymb,
 aps,
pre,
]{revtex4-1}

\RequirePackage{snapshot}
\usepackage[english]{babel}
\usepackage{amsfonts}
\usepackage{amsmath}
\usepackage{amssymb}
\usepackage{amsthm}
\usepackage{graphicx,psfrag}
\usepackage{color}
\usepackage{dirtytalk}
\usepackage{epigraph}
\usepackage{pxfonts}
\usepackage{stmaryrd}
\usepackage{listings}
\usepackage{diagbox}
\usepackage{makecell}
\usepackage{textcomp}
\usepackage{numprint}

\usepackage[pdftitle={Percolation Thresholds 
and Fisher Exponents 
in %the $d$-Dimensional Hypercube},
Hypercubic Lattices}, 
pdfauthor={Stephan Mertens and Cristopher Moore},
pdfcreator={},
pdfsubject={Percolation},
pdfproducer={},
pdfkeywords={Monte Carlo, percolation, thresholds},hyperfootnotes=false]{hyperref}
\usepackage{url}
\graphicspath{{../figures/}}

\IfFileExists{date.tex}{%
  \input{date.tex}

}

\newcommand*{\Z}{\mathbb{Z}}

\newcommand{\smax}{s_{\mathrm{max}}}

\newcommand{\pcbond}{p_c^{\text{bond}}}
\newcommand{\pcsite}{p_c^{\text{site}}}

\newcommand\T{\rule{0pt}{2.6ex}}       % Top strut
\newcommand\B{\rule[-1.2ex]{0pt}{0pt}} % Bottom strut

\theoremstyle{plain}

\begin{document}

\title{Percolation Thresholds 
and Fisher Exponents 
in %the $d$-Dimensional Hypercube},
Hypercubic Lattices}

\author{Stephan Mertens}
\email{mertens@ovgu.de}
\affiliation{Santa Fe Institute, 1399 Hyde Park Rd., Santa Fe, NM 87501, USA}
\affiliation{Institut f\"ur Physik, Universit\"at
  Magdeburg, Universit\"atsplatz~2, 39016~Magdeburg, Germany}

\author{Cristopher Moore}
\email{moore@santafe.edu}
\affiliation{Santa Fe Institute, 1399 Hyde Park Rd., Santa Fe, NM 87501, USA}

\date{\today}

\begin{abstract}
We use invasion percolation to compute highly accurate numerical values for bond and site percolation thresholds $p_c$ on the hypercubic lattice $\mathbb{Z}^d$ for $d = 4,\ldots,13$. We also compute the Fisher exponent $\tau$ governing the cluster size distribution at criticality. Our results support the claim that the mean-field value $\tau = 5/2$ holds for $d \ge 6$, with logarithmic corrections to power-law scaling at $d=6$.
 \end{abstract}

% 64.60.ah   Phase Transitions
% 02.70.-c   Computational techniques; simulations
% 02.70.Rr    General statistical methods
% 05.10.Ln   Monte Carlo methods

\maketitle

\section{Introduction}
\label{sec:intro}

According to Auguste Rodin, ``sculpture is the art of the hole and the
lump''~\cite{rodin}. Percolation is the \emph{science} of the hole and
the lump: each site of a lattice is considered a
lump with probability $p$ and a hole with probability $1-p$. Lumps that
are connected to each other form larger lumps, and in percolation
theory~\cite{grimmett:book,*stauffer:aharony:book} one studies the
structure of these large lumps as a function of $p$. 

Whereas the sculptor is confined to three-dimensional pieces of art, the scientist
can study structures in any dimension. In this contribution we 
study percolation on the hypercubic lattce $\Z^d$ in dimensions $d=4, \ldots,13$. 
In particular we present a method that allows us to approximate 
the critical density $p_c$ very efficiently. This is the density at which 
a lump---pardon, a cluster---first appears that spans the entire system.

For larger values of $d$, numerical simulations in $\mathbb{Z}^d$ are
challenging because the lattice quickly becomes too big to fit into
the memory of a computer. Grassberger~\cite{grassberger:03} avoided
this problem by growing single clusters at a given value of $p$ using
the Leath algorithm~\cite{leath:76a}. With this approach, he estimated  
the critical densities to 5 or 6 digits of accuracy. We will use another algorithm---invasion percolation---to get
even more accurate estimates of the critical densities. 

The paper is organized as follows.  We begin by explaining the basic invasion percolation algorithm and its efficient implementation for percolation on $\mathbb{Z}^d$.  In Section~\ref{sec:measuring-pc} we discuss how to compute $p_c$ from simple properties of the invasion cluster. Section~\ref{sec:results-pc} provides our numerical results, including our new estimates of $p_c$ and how they compare to previous results. In Sections~\ref{sec:tau} and~\ref{sec:log-corrections} we examine the scaling of the cluster size distribution, and to what extent it supports the claim that $d=6$ is the upper critical dimension.  We find that the Fisher exponent $\tau$ matches its mean-field value $5/2$ for $d \ge 6$, with logarithmic corrections at $d=6$.

\section{Invasion Percolation}
\label{sec:invasion}

Invasion percolation is a stochastic growth process that was introduced
as a model of fluid transport through porous
media~\cite{lenormand:bories:80,chandler:etal:82,wilkinson:willemsen:83}.
It starts with a single seed vertex of the underlying graph, and grows a cluster around it. 
There are two versions of the model which invade vertices or edges, 
which we use to study site and bond percolation respectively.  
For site percolation, we assign each vertex a random weight uniformly distributed
in the unit interval $[0,1]$. At each step, we add the neighboring vertex with the smallest
weight to the cluster, increasing the cluster size $N$ by one.
%yielding a cluster of mass $N=2$.  This process
%is iterated: at each step, we assign random weights to each previously
%unassigned vertex in the cluster's neighborhood, and add the
%neighboring vertex with the smallest weight to the cluster, increasing
%the mass $N$ by $1$. 
For bond percolation, we assign weights to edges
rather than vertices, and we extend the invasion cluster along the
edge incident to it with the smallest weight. For simplicity, we will focus our discussion here 
on site percolation.

The benefit of invasion percolation is that we do
not need to store a lattice large enough to hold the largest cluster
that we might encounter; for high-dimensional lattices this would be
computationally infeasible. Instead we only need to store the vertices
belonging to the cluster, and the weights of the neighboring sites, 
which we can choose ``on the fly'' as the cluster grows. 
Because the coordination number of the
lattice is fixed, the total number of vertices and weights we need to store
grows only linearly with the mass of the cluster.  

We use two data structures to keep track of the vertices
we have explored so far, and to find the boundary vertex with the smallest weight. 
A \emph{set} is used to hold all vertices that have already been assigned weights, 
and a \emph{priority queue}~\cite{dasgupta:algorithms} is used to hold the boundary, i.e., 
all vertices that have been assigned weights that are not already part of the cluster. 
The priority queue lets us select and remove the lowest-weight vertex from the boundary, or add new vertices to
it, in time logarithmic in the size of the boundary. The set lets us add, remove, or search for a vertex in a cluster of size $N$ in time $O(\log N)$. Thus each step of invasion percolation takes $O(\log N)$ time.

Modern programming languages provide built-in implementations of these data structures, such as the 
container classes \texttt{set} and \texttt{priority\_queue} from the C++ standard library~\cite{vanweert:gregoire:16}. While these implementations are convenient, we found it more efficient to build our own implementation of the \texttt{set} data structure as a hash table, using essentially the same hash function as in~\cite{grassberger:03}.
%, see Appendix~\ref{sec:implementation}.

% In order to achieve the logarithmic time complexity of the container
% classes, the vertices have to be sortable. The vertices in the
% $d$-dimensional lattices are naturally ordered according to the
% lexicographic order of their coordinates.

\emph{A priori}, invasion percolation differs from classical Bernoulli
percolation, where each vertex is independently occupied with
probability $p$.  But invasion percolation reproduces, both
qualitatively and quantitatively, Bernoulli percolation at
criticality~\cite{chayes:chayes:newman:85,haeggstroem:peres:schonmann:99}.
We can explain this connection as follows.  Since the
vertex weights are uniform in the unit interval, one way to implement
Bernoulli percolation is to declare a vertex occupied if its weight is
less than $p$.  If $p > p_c$, the occupied sites possess a unique infinite component. 
If the initial vertex is in a finite component, invasion percolation fills it and then breaks out of it by 
adding a vertex with weight greater than $p$; but after an initial transient of these finite components, 
it breaks through to the infinite component. After that, it grows the cluster to infinite mass 
by adding vertices of weight less than $p$.  

Thus, for any $p > p_c$, the maximum weight of the vertices added to the cluster falls below $p$ 
after some finite time. As the cluster grows, the maximum weight approaches $p_c$ from above, 
so the weights of the vertices added to the cluster are asymptotically uniform in the 
interval $[0,p_c]$.

\section{Measuring the Threshold}
\label{sec:measuring-pc}

To compute $p_c$ using invasion percolation, we use a simple estimator. Let $B(N)$ denote the number of vertices that have been assigned weights in the course of building a cluster with mass $N$, i.e., which are either in the cluster or on its boundary. Since almost all of the $N$ vertices actually added to the cluster have weight less than or equal to $p_c$, and since the weight distribution is uniform, we have
\begin{equation}
  \label{eq:volume-surface-ratio}
  \lim_{N \to \infty} \frac{N}{B(N)} = p_c \, .
\end{equation}
The relationship between $p_c$ and this ``boundary-to-bulk ratio'' goes back to the classic work of Leath~\cite{leath:76a}. The limit~\eqref{eq:volume-surface-ratio} has been established rigorously for invasion bond percolation in $\Z^2$~\cite{chayes:chayes:newman:85}, but it is believed to hold on general lattices. Moreover, it appears to be an excellent numerical estimator for $p_c$, and in~\cite{mertens:moore:17} we used it to estimate $p_c$ to high precision on hyperbolic lattices. 

Based on these facts, we estimate $p_c$ by extrapolating the measured values $N/B(N)$ to $N=\infty$.  The estimator $N/B(N)$ is extremely easy to compute, since $N$ and $B(N)$ are simply integers given by the progress of the invasion percolation process. Moreover, it turns out to have excellent finite-size scaling and small statistical fluctuations. 

We first consider the behavior of this estimator on a Bethe lattice of degree $\Delta$, i.e., a tree where every vertex has $\Delta-1$ children. Here we have $p_c = 1/(\Delta-1)$ for both site and bond percolation. Moreover, any cluster of mass $N$ is surrounded by exactly $(\Delta-2)N+2$ neighboring vertices regardless of its shape, giving
\begin{equation}
  \label{eq:volume-surface-ratio-tree}
  \frac{N}{B(N)} 
  = \frac{N}{(\Delta-1)N+2} 
  = \frac{p_c}{1+\frac{2}{(\Delta-1)N}} 
 \, .
\end{equation}
%Thus the value of $N/B(N)$ does not depend on the
%random weights chosen to grow the invasion cluster, or on the shape of the tree: it is a
%deterministic quantity, which
Thus on the tree this estimator has zero variance, and a finite-size effect which decays as $O(N^{-\delta})$ for $\delta=1$.

We find that the convergence of this estimator is almost as good on $d$-dimensional lattices as it is on the tree. As in~\cite{mertens:moore:17} we assume the form
\begin{equation}
  \label{eq:finite-size-scaling}
  \frac{N}{B(N)} = \frac{p_c}{1 + b N^{-\delta}} \simeq p_c(1-b N^{-\delta})
\end{equation}
and fit the parameters $p_c$, $b$, and $\delta$ to our numerical data. We will see that $\delta$ quickly approaches $1$  as $d$ increases. This is plausible since high-dimensional lattices are treelike, in the sense that most paths never return to the origin. Moreover, $N/B(N)$ has small statistical fluctuations, so we can obtain accurate estimates of $p_c$ with a reasonable number of independent runs of the algorithm.

It is interesting to consider the relationship between $\delta$ and the transient described above. The invasion cluster of mass $N$ contains both ``good'' vertices, i.e., those with weights in the interval $[0,p_c]$, and ``breakout'' vertices of weight greater than $p_c$, which are added to the cluster whenever it fills a finite component and runs out of good vertices on the boundary. One might think that the error in the estimator comes from these breakout vertices: for instance, one might think that if there are $N_b = O(N^\alpha)$ breakout vertices in the cluster, then the error is $O(N_b/N) = O(N^{-\delta})$ where $\delta = 1-\alpha$.

However, this is overly pessimistic. On the tree, each boundary vertex has $\Delta-1$ children; the probability that a given child is good is $p_c=1/(\Delta-1)$, so the expected number of good children is $1$. Since invasion percolation will add a good boundary vertex to the cluster if there is one, it follows that the number $N_g$ of good boundary vertices follows an unbiased random walk. A breakout occurs when $N_g=0$, i.e., when this random walk touches the origin. Since an unbiased random walk touches the origin $O(N^{1/2})$ times in its first $N$ steps, the cluster includes $N_b=O(N^{1/2})$ breakout vertices. On the other hand, after $N$ steps there are also typically $N_g=O(N^{1/2})$ good vertices waiting to be added to the cluster. Miraculously, these two cancel out, and $N = p_c B(N)+O(1)$ as shown in~\eqref{eq:volume-surface-ratio-tree}. We suspect that a similar cancellation occurs on lattices, so that $\delta$ is smaller than one would expect from naively considering the number of breakouts before the infinite component is reached.

\section{Critical Densities}
\label{sec:results-pc}

\begin{figure}
  \centering
  \includegraphics[width=\columnwidth]{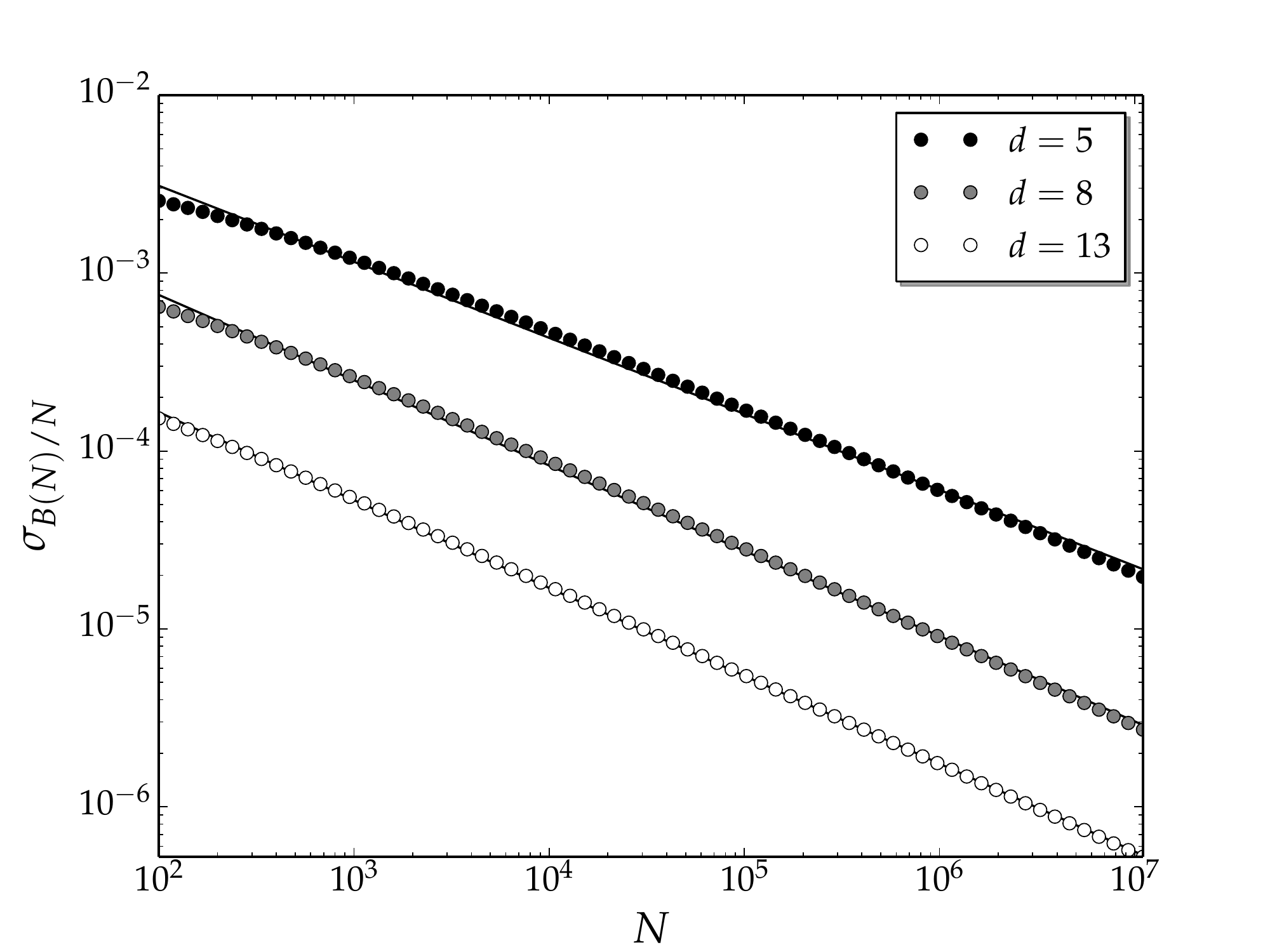}
  \caption{The standard deviation of $B(N)/N$ vs.\ $N$ over multiple runs of invasion percolation. The straight lines show that the statistical fluctuations decay as $O(N^{-1/2})$ for large $d$. For $d=5$ the decay is slightly slower, $O(N^{-0.46})$.}
  \label{fig:sigma_B}
\end{figure}

To compute $p_c$ numerically, we grow an invasion cluster and report values of $N/B(N)$ for  
\begin{equation}
  \label{eq:N-sampling}
  N = \lfloor 100 \times 2^{t/4} \rfloor \qquad (t=0,1,\ldots,67) \, ,
\end{equation}
where $\lfloor x \rfloor$ denotes the integer part of $x$. This correponds to cluster sizes ranging from $100$ to $\numprint{11021797}\approx 10^7$. These $68$ data points are then averaged over $10^6$ independent runs of the invasion percolation algorithm. Figure~\ref{fig:sigma_B} shows that the statistical fluctuations between runs are small: for large $d$ the standard deviation decays as $O(N^{-1/2})$, just as if the $N$ steps of invasion percolation were independent events. 

\begin{figure}
  \centering
  \includegraphics[width=\columnwidth]{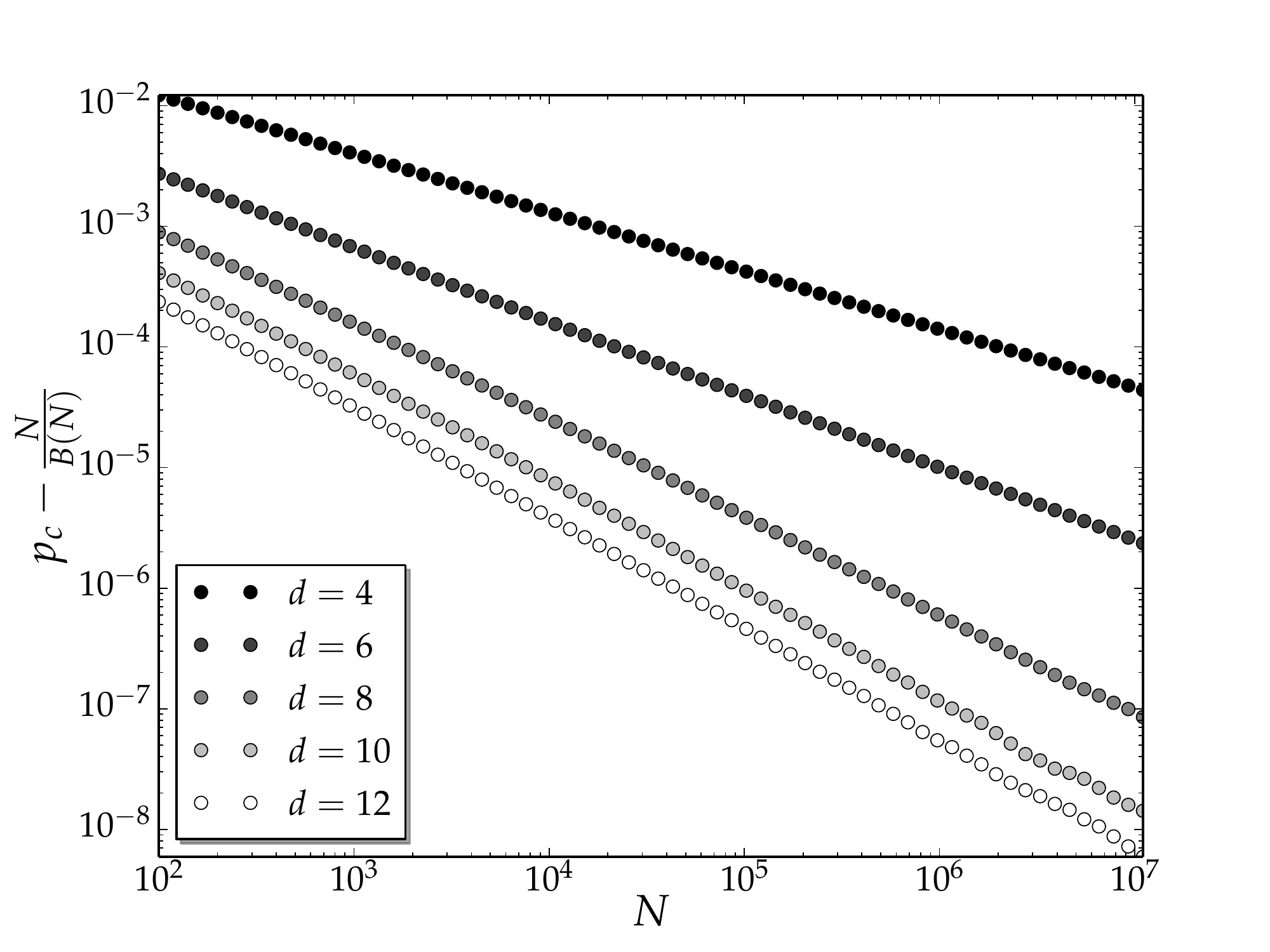}
  \caption{Convergence of $N/B(N)$ to $\pcsite$. The finite-size effects decay as $O(N^{-\delta})$, where $\delta$ approaches $1$ for large $d$.}
  \label{fig:pc-delta}
\end{figure}

The data follows~\eqref{eq:finite-size-scaling} very well, as can be seen from Figure~\ref{fig:pc-delta}. Fitting $p_c$, $b$, and $\delta$, and thus extrapolating $N/B(N)$ to $N \to \infty$, then allows us to estimate $p_c$. Table~\ref{tab:delta} shows that the exponent $\delta$ governing the finite-size effects increases with $d$, approaching its treelike value $1$. Together with the small standard deviation of $B(N)/N$, these values of $N$ and the sample size of $10^6$ are large enough to estimate $p_c$ quite precisely.

\begin{table}
\centering
$
\begin{array}{ll}
d & \delta \\ \hline
4 & 0.483\,21(6) \\
5 & 0.518\,3(3) \\
6 & 0.601\,2(3) \\
7 & 0.717\,9(4) \\
8 & 0.812(1) \\
9 & 0.870(2) \\
10 & 0.903(2) \\
11 & 0.911(1) \\
12 & 0.916(1) \\
13 & 0.935(1)
\end{array}
$
\caption{Numerical values of the exponent $\delta$ governing the finite-size effects~\eqref{eq:finite-size-scaling}. As $d$ increases, $\delta$ approaches its value $1$ on the tree.\label{tab:delta}} 
\end{table}

Table~\ref{tab:hypercube} contains the values for $p_c$ obtained by our approach, for both site and bond percolation, in comparison to the most precise values we could find in the literature. Our values for $p_c$ are consistent with the previous results of~\cite{grassberger:03,dammer:hinrichsen:04} but are more accurate. Note that, because $\delta$ increases with $d$, our estimates become more accurate as $d$ increases:  
%the finite size effects decay faster in higher dimensions, which is plausible since in higher dimensions, the lattice becomes more tree like. 
with invasion clusters of size $N=10^7$, our multiplicative error on $p_c$ is $10^{-4}$ for $d=4$, but just $10^{-8}$ for $d=13$. Our new values have error bars that are at least a factor $10^{-2}$ smaller than those of the previous values, giving $p_c$ to two or three more digits of accuracy. 

\begin{table}
  \centering
  \begin{tabular}{r@{\hskip 12pt}l@{\hskip 12pt}l}
     $d$\T\B & \multicolumn{1}{c}{$\pcsite$} &
                                               \multicolumn{1}{c}{$\pcbond$} \\\hline
      4 & 0.196\,886\,1(14)\textsuperscript{\cite{grassberger:03}}\T & 0.160\,131\,0(10)\textsuperscript{\cite{dammer:hinrichsen:04}}\\
         & 0.196\,885\,61(3) & 0.160\,131\,22(6)\\
      5 & 0.140\,796\,6(15)\textsuperscript{\cite{grassberger:03}}\T & 0.118\,171\,8(3)\textsuperscript{\cite{dammer:hinrichsen:04}}\\
         & 0.140\,796\,33(4) & 0.118\,171\,45(3)\\
      6 & 0.109\,017(2)\textsuperscript{\cite{grassberger:03}}\T & 0.094\,201\,9(6)\textsuperscript{\cite{grassberger:03}}\\
         & 0.109\,016\,661(8) & 0.094\,201\,65(2)\\
      7 & 0.088\,951\,1(9)\textsuperscript{\cite{grassberger:03}}\T & 0.078\,675\,2(3)\textsuperscript{\cite{grassberger:03}}\\
         & 0.088\,951\,121(1) & 0.078\,675\,230(2) \\
      8 & 0.075\,210\,1(5)\textsuperscript{\cite{grassberger:03}}\T & 0.067\,708\,39(7)\textsuperscript{\cite{grassberger:03}}\\
         & 0.075\,210\,128(1) & 0.067\,708\,418\,1(3)\\
      9 & 0.065\,209\,5(3)\textsuperscript{\cite{grassberger:03}}\T & 0.059\,496\,01(5)\textsuperscript{\cite{grassberger:03}}\\
         & 0.065\,209\,5348(6) & 0.059\,496\,003\,4(1)\\
    10 & 0.057\,593\,0(1)\textsuperscript{\cite{grassberger:03}}\T & 0.053\,092\,58(4)\textsuperscript{\cite{grassberger:03}}\\
         & 0.057\,592\,948\,8(4) & 0.053\,092\,584\,2(2)\\
    11 & 0.051\,589\,71(8)\textsuperscript{\cite{grassberger:03}}\T & 0.047\,949\,69(1)\textsuperscript{\cite{grassberger:03}}\\
         & 0.051\,589\,684\,3(2) & 0.047\,949\,683\,73(8)\\
    12 & 0.046\,730\,99(6)\textsuperscript{\cite{grassberger:03}}\T & 0.043\,723\,86(1)\textsuperscript{\cite{grassberger:03}}\\
         & 0.046\,730\,975\,5(1) & 0.043\,723\,858\,25(10)\\
    13 & 0.042\,715\,08(8)\textsuperscript{\cite{grassberger:03}}\T &
                                                                  0.040\,187\,62(1)\textsuperscript{\cite{grassberger:03}}\\
         & 0.042\,715\,079\,60(10) & 0.040\,187\,617\,03(6)
  \end{tabular}
  \caption{Previous and new numerical values for the percolation thresholds on the
    $d$-dimensional hypercube. }
  \label{tab:hypercube}
\end{table}

%\textbf{{\color{red}TO DO:}} 
%Compare values of $p_c$ to upper bounds on $p_c$ from
%Pad\'e approximants of the mean cluster size \cite{torquato:jiao:13a}.
%And from the series 
%series expansion for the reciprocal of the connective constant in
%$\mathbb{Z}^d$ \cite{hara:slade:95}: 
%\begin{equation}
%  \label{eq:hara-slade}
%  \frac{1}{\mu(d)} = \sigma^{-1} + \sigma^{-3}+2\sigma^{-4}+12\sigma^{-5}+66\sigma^{-6}+O(\sigma^{-7}) \, .
%\end{equation}
%with
%\begin{equation}
%  \label{eq:def-sigma}
%  \sigma = 2d-1
%\end{equation}
%We could also compare it to ``our'' new series for $p_c$, copying some
%of the discussions from the other paper.

%Other $d$-dimensional lattices: Kagom\'e lattice \cite{vandermarck:98},
%diamond, fcc and bcc lattices \cite{vandermarck:98a}

\section{Fisher Exponent $\mathbf{\tau}$}
\label{sec:tau}

Now that we have precise values for $p_c$, we can measure other quantities at criticality, such as the cluster size distribution and critical exponents.  According to scaling theory, the average number of clusters of size $s$ per lattice site scales as
\begin{equation}
  \label{eq:ns-scaling}
  n_s(p) = s^{-\tau}\,(f_0(z) + s^{-\Omega} f_1(z) + \cdots) \, ,
\end{equation} 
where the exponent $\Omega$ governs the leading finite-size corrections (e.g.~\cite{ziff:11}).  The scaling functions $f_0$ and $f_1$ are analytic for small values of $z$, and where the scaling variable $z$ is defined as
\begin{equation}
  \label{eq:z-scaling}
  z = (p-p_c) s^\sigma \, .
\end{equation}
(We ignore nonuniversal metric factors in $z$.)  
At $p=p_c$ we have $z=0$, and~\eqref{eq:ns-scaling} becomes
\begin{equation}
  \label{eq:ns-scaling-pc}
n_s(p_c) = s^{-\tau} \,(c_0 + c_1 s^{-\Omega} + \cdots ) \, ,
\end{equation}
for some constants $c_0 = f_0(0)$ and $c_1 = f_1(0)$. 

The critical exponents $\tau$ and $\sigma$ are known exactly for percolation in two dimensions from conformal field theory and exactly solvable models (e.g.~\cite{nienhuis:riedel:schick:80}), and on the Bethe lattice from mean field theory:
\begin{equation}
  \label{eq:tau-sigma-exact}
  (\tau,\sigma) = \begin{cases}
      \left(\frac{187}{91}, \frac{36}{91}\right) & d=2 \, , \\
      \left(\frac{5}{2}, \frac{1}{2}\right) & \mbox{Bethe lattice.}
  \end{cases}
\end{equation}
It is believed that upper critical dimension for percolation is $d_c = 6$, i.e., that $\tau$ and $\sigma$ take their mean-field values for $d \ge d_c$. The original argument by Toulouse~\cite{toulouse:74} rests on the Josephson hyper-scaling relation
\begin{equation}
  \label{eq:josephson}
  d \nu = 2\beta + \gamma \, ,
\end{equation}
that relates the critical exponents $\nu$, $\beta$, and $\gamma$ with the spatial dimension $d$.  For the mean-field solution on the Bethe lattice we have $\beta = \gamma = 1$ and $\nu=\frac{1}{2}$, which implies $d_c=6$. Rigorous results provide only upper bounds for the critical dimension; the best known bounds are $6 \le d_c \le 10$~\cite{chayes:chayes:87,fitzner:vanderhofstad:17}.  We will compute $\tau$ numerically for $3 \le d \le 10$ and see that these values support the claim that $d_c=6$, with logarithmic corrections at $d=d_c$.

To measure the critical exponent $\tau$, we grow a cluster outward from the origin using the Leath algorithm~\cite{leath:76a,*leath:76b}. This is similar to invasion percolation except that it adds every neighboring vertex with weight $p \le p_c$ to the cluster, instead of just the vertex with minimum weight. 
%It is straightforward to modify the invasion percolation algorithm and our data structures appropriately (see Appendix). 
By running the Leath algorithm $T$ times at $p=p_c$, we generate $T$ independent samples from the cluster size distribution at criticality. The probability $P(s)$ that a given occupied vertex is contained in a cluster of size $s$ scales as $sn_s \sim s^{-(\tau-1)}$, so we can infer $\tau$ from the fraction $T_s/T$ of runs in which the Leath algorithm gives a cluster of each size $s$. 

We fit $\tau$ to the data from these experiments using the complementary cumulative distribution $Q(s) = \sum_{s'=s}^\infty P(s')$. Including the term of~\eqref{eq:ns-scaling-pc} governing the leading finite-size effects and approximating this sum as an integral gives the form
\[
Q(s) = C s^{-(\tau-2)} (1 + a s^{-\Omega}) 
\, , 
\]
for some constants $C$ and $a$, and taking the logarithm gives
\begin{align}
\log Q(s) &= -(\tau-2) \log s + \log(1 + a s^{-\Omega}) + c 
\label{eq:q1} \\
&\approx -(\tau-2) \log s + a s^{-\Omega} + c 
\label{eq:q2}
\end{align}
where $c=\log C$. We find the parameters $\tau$, $\Omega$, $a$, and $c$ using a nonlinear least-squares fit to $Q(s)$ of this form; we found that using~\eqref{eq:q1} vs.\ \eqref{eq:q2} changes our estimate of $\tau$ only very slightly, within the error bars we report below. We also found that representing the sum $\sum_{s'=s}^\infty {s'}^{-\alpha}$ exactly with the Hurwitz zeta function $\zeta(\alpha,s)$ makes no appreciable difference to our results. 

% may need to be updated 
For each experiment we generate $T$ clusters, where in most cases $T=10^8$. For site percolation we did more extensive experiments near the critical dimension, with $T=10^9$ for $d=5, 6, 7$. We reduce our computation time by stopping the Leath algorithm at a maximum size $\smax=2^{20}$, counting the clusters that reach this size toward $Q(s)$ for all $s \le \smax$. 
%We took $\smax = 2^{20}$ for $d=3,4$ and $\smax=2^{23}$ for $5 \le d \le 10$. For bond percolation we took $T=10^8$ and $\smax = 2^{20}$ for all $d$.
We bin the data logarithmically, computing $Q(s)$ for $s=\lfloor 2^{k/4} \rfloor$ for integer $k$. 
%For site percolation 
We discard the first $30$ bins, i.e., clusters of size $180$ or less. 
(Note that for bond percolation we define the size of a cluster as the number of edges it contains.) 
%; for bond percolation we ignore the first $30$ bins, i.e., clusters containing $180$ or fewer edges. 

In order to estimate the error bars on our estimate of $\tau$, we perform a nonparametric bootstrap~\cite{efron:gong:83,shalizi:10}, resampling $T$ cluster sizes with replacement from the original empirical distribution. We perform $1000$ independent trials of this bootstrap procedure, fit $\tau$ to each one, and define the 90\% confidence interval by cutting off the lowest 5\% and the highest 5\% values of $\tau$. This gives the estimates and error bars shown in Table~\ref{tab:tau}.

We can estimate the finite-size scaling exponent $\Omega$ using the same procedure, albeit with much less precision than $\tau$. For $d=3$ we obtained $\Omega=0.77(3)$, which is larger than the value $0.63(2)$ reported in~\cite{lorenz:ziff:98}. However, we found that our estimate of $\Omega$ depends on how many initial bins we discard, suggesting that there are competing finite-size effects that mask the first term in~\eqref{eq:ns-scaling-pc}.

\begin{table}
  \centering
  $
%  \begin{tabular}{c@{\hskip 12pt}ll}
%    d & \multicolumn{1}{c}{\tau^{\mathrm{site}}} & \multicolumn{1}{c}{\tau^{\mathrm{bond}}} \\\hline
%     3 & 2.18922(2)\T & 2.1897(2) \\ % 2.189(1)
%     4 & 2.3114(3) \T & 2.3148(2)\\ % 2.314(1)
%     5 & 2.4096(3) \T & 2.4160(6)\\ % 2.414(2)
%     6 & 2.4667(2) \T & 2.4759(7) \\ % 2.479(2)
%     7 & 2.4869(9) \T & 2.4955(8)\\ % 2.500(4)
%     8 & 2.4972(3) \T & 2.4990(6)\\ % 2.501(3)
%     9 & 2.4984(1) \T & 2.4996(7)\\ % 2.503(4)
\begin{array}{cll}
d
& \tau^{\mathrm{site}} 
%& \tau^{\mathrm{site}} (MLE) 
& \tau^{\mathrm{bond}} 
%& \tau^{\mathrm{bond}} (MLE)
%& \;\;\; \Omega
 \\ \hline
      3 
      & 2.1892(1)
%      & 2.1891(3) 
      & 2.1890(2)
%      & 2.1890(3) 
\\
%      0.65(3) \\
     4 
     & 2.3142(5)
%     & 2.3140(8) 
     & 2.311(2)
%     & 2.313(1) 
\\
%     0.37(3) \\
     5 
     & 2.419(1)
%     & 2.416(1) 
     & 2.422(4)
%     & 2.418(1) 
     \\
     6 
     & 2.487(2)^*
%     & 2.477(3)^* 
     & 2.488(6)^*
%     & 2.483(3)^* 
     \\
     7 
     & 2.499(1) 
%     & 2.496(3) 
     & 2.499(2)
%     & 2.498(3) 
     \\
     8 
     & 2.499(2)
%     & 2.501(4) 
     & 2.500(1)
%     & 2.498(3) 
     \\
     9 
     & 2.501(2)
%     & 2.503(4) 
     & 2.505(5)
%     & 2.501(3) 
     \\
     10 
     & 2.503(4)
%     & 2.497(4) 
& 2.498(2) 
\end{array}
$
%  \end{tabular}
  \caption{The Fisher exponent $\tau$ for percolation on the $d$-dimensional hypercubic lattice, measured using a least-squares fit to a power law with finite-size corrections of the form~\eqref{eq:q1}.  Error bars shown are $90\%$ confidence intervals obtained by bootstrap resampling. For $d > 6$ these results are consistent with the mean-field value $5/2$. At the critical dimension $d=6$ the measured exponent (marked with $*$) is depressed due to logarithmic corrections; see text.}
  \label{tab:tau}
\end{table}

For $d=3$ our results are consistent with the recent value $\tau = 2.18909(5)$~\cite{xu:etal:14}. For $d=4$ our value is consistent with the previous result $\tau = 2.313(2)$~\cite{tiggemann:01,paul:ziff:stanley:01}, but (for site percolation) with smaller error bars. For $d=5$ our estimate is larger than the previous value $\tau=2.412(4)$ from~\cite{paul:ziff:stanley:01}, which we think is due to larger cluster sizes and better avoidance of finite-size effects. Our results are also quite close to  the estimates $2.1888$ ($d=3$), $2.3124$ ($d=4$), and $2.4171$ ($d=5$)  from four-loop renormalization theory~\cite{gracey:15}.  

For $d > 6$, our values are consistent with the mean-field prediction $\tau = 5/2$.  This result was also recently obtained for $d=7$ in~\cite{huang:etal:18}. However, for $d=6$, a fit of the form~\eqref{eq:q1} yields an estimate of $\tau$ significantly below its mean-field value $5/2$. This is due to the fact that the leading term is modified by logarithmic corrections, as we discuss in the next section.

%We can also estimate the finite-size scaling exponent $\Omega$ in~\eqref{eq:ns-scaling} by observing how our estimates of $\tau$ vary with $\smin$. Specifically, the estimate $\hat{\tau}$ behaves as $|\hat{\tau} - \tau| = O(\smin^{-\Omega})$.  For $d=3$ this gives $\Omega=0.65(3)$, consistent with the value $0.63(2)$ found in~\cite{lorenz:ziff:98}. 

\section{Logarithmic Corrections at $d=6$}
\label{sec:log-corrections}

At the critical dimension $d_c=6$, we expect the size distribution of clusters to obey mean-field theory but with logarithmic corrections~\cite{essam:etal:78,nakanishi:stanley:80},
\[
n_s \sim s^{-\tau} (\log s)^\theta \, .
\]
To be more specific, we expect the cumulative size distribution of the cluster containing the origin to scale as
\[
Q(s) \sim s^{-(\tau-2)} (\log s)^\theta \, ,
\]
or
\begin{equation}
\label{eq:q-log}
\log Q(s) = -(\tau-2) \log s + \theta \log \log s + c
\end{equation}
for some constant $c$.

Using the same data as in the previous section, namely $10^9$ site percolation clusters grown at $p=p_c$ using the Leath algorithm up to a maximum size $\smax=2^{20}$, with sizes binned logarithmically, we find $\tau$, $\theta$, and $c$ using a nonlinear least-squares fit to the form~\eqref{eq:q-log}. We again obtain a 90\% confidence interval using the bootstrap, fitting these parameters to 1000 independently resampled data sets. To avoid finite-size effects we discard the first $40$ bins, i.e., clusters of size less than $1024$. 

Using this approach, we obtain $\tau = 2.501(1)$, consistent with the mean-field value $5/2$. Our estimate for $\theta$ is $0.34(1)$.  This is inconsistent with the prediction $\theta=2/7$ from renormalization group techniques~\cite{essam:etal:78,nakanishi:stanley:80}. However, as with the finite-size exponent $\Omega$ for $d=3$, our estimate of $\theta$ is sensitive to how many initial bins we discard, presumably because it is confounded by finite-size effects. If we discard the first $60$ bins, ignoring clusters of size less than $2^{15} = 32\,768$, we obtain $\tau = 2.498(5)$ and $\theta=0.32(6)$, consistent with theory. We hope to provide more precise measurements of $\theta$ with additional numerical work in the future.

\bigskip
\section{Conclusions}

We implemented invasion percolation using efficient data structures on the hypercubic lattice, and used a simple, rapidly-converging estimator based on the boundary-to-bulk ratio to obtain highly accurate measurements of the critical densities $p_c$ for site and bond percolation. This approach has the benefit that we do not need to perform multiple runs at different values of $p$, nor do we need to choose a criterion for criticality such as clusters that cross or wrap around a finite lattice. Instead, the estimate of $p_c$ emerges naturally from the process. Moreover, by storing the invasion cluster in a hash table, we can grow much larger clusters than we could store in memory in a surrounding lattice.

Using the size distribution of clusters at $p=p_c$ using the Leath algorithm, we computed numerical values of the Fisher exponent $\tau$, confirming that its mean-field value $\tau=5/2$ holds for $d \ge 6$ but with logarithmic corrections at the critical dimension $d=6$. There are many other critical exponents that one could measure with this approach, as well as quantities such as the ratio of the mean cluster size above and below $p_c$. We leave these for future work.

It is interesting to consider measuring critical behavior using invasion percolation directly, rather than by first estimating $p_c$ and then using the Leath algorithm. The invasion cluster consists of a union of connected clusters at $p=p_c$, so as it grows its statistical properties approach those of the giant cluster that appears at the transition. This suggests that its surface area, radius of gyration, and other properties scale like those of the critical cluster in the limit $N \to \infty$, and indeed in two and three dimensions the invasion cluster and critical cluster have the same fractal dimension~\cite{wilkinson:willemsen:83,zhang:95}. In the same spirit as other notions of ``self-organized percolation''~\cite{alencar:andrade:lucena:97,grassberger:zhang:96}, we hope invasion percolation will allow us to measure other properties of the critical cluster without tuning the parameter $p$. We leave this for future work as well.

%\appendix
%
%\section{Implementation and Technical Details}
%\label{sec:implementation}
%
%\begin{table}
%\bigskip
%  \centering
%  \begin{tabular}{lccc}
%  CPU & frequency & nodes$\times$cpus$\times$cores & memory/core \\[0.5ex]
%  E5-1620 & 3.60 GHz & $1\times 2\times 4$ & 4.0 GByte \\
%  E5-2630 & 2.30 GHz & $5\times 4\times 6$  & 5.3 GByte \\
%  E5-2630v2 & 2.60 GHz &  $5\times 4\times 6$ & 5.3 GByte \\
%  E5-2640v4 & 2.40 Ghz & $3\times 4 \times 10$ & 6.4 GByte
%  \end{tabular}
%  \caption{Computing machinery used for the simulations in this
%    paper. All CPUs are Intel\textsuperscript{\textregistered} Xeon\textsuperscript{\textregistered}.}
%  \label{tab:leonardo}
%\end{table}
%
%We ran our simulations on a cluster with a mixture of CPUs and a total of 368 cores, see Table~\ref{tab:leonardo}. Each value of Table~\ref{tab:hypercube} took roughly \todo{??} hours wall-clock time on this cluster. The actual invasion percolation algorithm has time complexity $O(dN\log(dN))$ for each run.

%\paragraph*{Acknowledgments.} 
\acknowledgements{
S.M. thanks the Santa Fe Institute for their hospitality, and C.M. thanks Doro Frederking for hers. We are grateful to Bob Ziff and Cosma Shalizi for helpful conversations.
}

\bibliography{percolation,animals,statmech,mertens,cs,math}

\end{document}